\documentclass[conference]{IEEEtran}

\usepackage{amsmath,amssymb,amsfonts}

\usepackage{cite}

\usepackage[table]{xcolor}
\definecolor{fcblue}{RGB}{70,130,180}
\definecolor{zblue}{RGB}{191,217,245}
\definecolor{tgreen}{RGB}{72,163,7}
\definecolor{zgreen}{RGB}{202,247,169}
\definecolor{hmblue}{HTML}{185FA5}

\usepackage{graphicx}
\usepackage{tikz}
\usetikzlibrary{positioning,arrows.meta,fit,calc}
\usepackage{url}
\usepackage{booktabs}
\usepackage{tabularx}
\usepackage{array}
\usepackage{makecell}
\usepackage{multirow}

\definecolor{fcblue}{RGB}{70,130,180}
\definecolor{zblue}{RGB}{191,217,245}
\definecolor{tgreen}{RGB}{72,163,7}
\definecolor{zgreen}{RGB}{202,247,169}
\definecolor{hmblue}{HTML}{185FA5}

\usepackage{algorithmic}
\usepackage{textcomp}
\usepackage[normalem]{ulem}

\def\BibTeX{{\rm B\kern-.05em{\sc i\kern-.025em b}\kern-.08em
    T\kern-.1667em\lower.7ex\hbox{E}\kern-.125emX}}

\begin{document}

\title{Surface Cues Over Structure: Probing LLM Behavior on
Brain Connectivity Graphs}

\author{
\IEEEauthorblockN{Prapti Roy and Khaled Mohammed Saifuddin}
\IEEEauthorblockA{
School of Computing, Southern Illinois University, Carbondale, IL, USA
}
}

\maketitle

\begin{abstract}
Resting-state functional magnetic resonance imaging (rs-fMRI) provides a noninvasive view of functional interactions across the brain, which can be represented as connectivity graphs for neurological disease analysis. Large language models (LLMs) are increasingly being paired with such representations, yet it remains unclear whether they genuinely reason over the underlying connectivity or simply respond to surface-level cues that may trigger learned priors. We probe this question through a controlled study of Attention-Deficit/Hyperactivity Disorder (ADHD) diagnosis from resting-state functional connectivity, isolating one factor at a time to determine what drives an LLM's predictions on brain connectivity graphs. Across six models, from small open-weight LLMs to a proprietary one, the same pattern holds: the results provide little evidence that predictions are consistently grounded in subject-specific connectivity. Brain region names, numeric formatting, and prompt framing shift what a model predicts without improving how well it separates ADHD subjects from Controls; perturbing the graph changes predictions no more for clinically meaningful regions than for random ones; and the models' own justifications contradict the values they cite, inconsistently interpreting the connectivity evidence provided to them. Across every probe, predictions follow the models' built-in class bias and the surface form of the prompt rather than the subject's connectivity, calling into question a premise underlying current LLM–brain integration work. Our code is available at \url{https://github.com/Roy025/ADHD-code}.
\end{abstract}

\begin{IEEEkeywords} brain network analysis, large language model, functional connectivity
\end{IEEEkeywords}

\section{Introduction}
\label{sec:intro}
The brain operates as a network\cite{bullmore2009complex}. Neurological disorders such as Attention-Deficit/Hyperactivity Disorder (ADHD), autism, and Alzheimer's disease manifest not as dysfunction in isolated brain regions but in disrupted patterns of connectivity between them~\cite{wang2009altered,farahani2019application}. Resting-state functional magnetic resonance imaging (rs-fMRI) captures these patterns by tracking correlated activity across brain regions at rest. Because the disruptions of interest lie in the connections between regions, these signals are naturally represented as functional connectivity (FC) graphs, in which nodes are regions and edges encode their pairwise functional synchrony~\cite{rubinov2010complex, biswal1995functional}.
As large language models (LLMs) have demonstrated sophisticated reasoning across domains\cite{truhn2023large, luo2025large, wei2022chain, singhal2025toward} a natural question has emerged: \textit{Can LLMs reason directly over brain connectivity graphs, identifying clinically meaningful structural patterns and supporting diagnostic prediction?}

A growing body of work has begun to evaluate LLM reasoning on graphs, testing whether models can determine node degree, connectivity, and other structural properties when graphs are represented as text \cite{guo2023gpt4graph, tang2025grapharena, fatemi2024talk, dai2024revisiting, zhang2024llm4dyg}. Studies show that encoding choices, particularly whether nodes carry meaningful names or anonymous identifiers, substantially affect model behavior~\cite{fatemi2024talk, dai2024revisiting, zhang2024llm4dyg}. However, they do not test this on biomedical graphs, particularly brain connectivity graphs.
Separately, a distinct line of work incorporates LLMs into brain graph analysis. These methods use an LLM to generate an intermediate connectivity representation that is then passed to a downstream task, such as distinguishing individuals with a disorder from healthy Controls~\cite{dong2026bleg,wei2026fmri}. Yet a basic question remains untested: whether anatomical regions of interest (ROIs), such as \textit{Frontal\_Sup\_L} or \textit{Caudate\_L} serve as neutral identifiers carrying no more information than arbitrary node indices, or instead inject disorder-relevant pretraining associations, and whether LLM outputs reflect reasoning over connectivity structure at all.

This premise is far from obvious. LLMs encode substantial relational and factual knowledge acquired during pretraining, much of which can be retrieved without task-specific fine-tuning~\cite{petroni2019language,wang2024unveiling}. Brain regions are unlikely to be an exception, as associations between anatomical regions and neurological or psychiatric disorders are extensively documented in the biomedical literature~\cite{menon2011large, farahani2019application}. Such associations may also be reflected in LLMs' pretrained knowledge through exposure to large-scale scientific and biomedical text~\cite{gao2020pile,usuyama2025biomedical}.

Such knowledge may be partly illusory, as factual-query performance can be driven by surface forms of entity names rather than by retrieval of the underlying facts~\cite{poerner2020bert, nishida2026revisiting}.
When an LLM sees \textit{Frontal\_Sup\_L} in a brain graph, it may not read the connectivity values at all; it may simply recall that frontal connectivity is disrupted in ADHD and respond accordingly, regardless of what the edge weights indicate. We call this the \textbf{surface cue hypothesis}: that an LLM's predictions on brain graphs are shaped by surface-level cues, such as anatomical names, numeric formatting, and prompt framing, rather than by the connectivity structure, and that responding to these cues does not amount to genuine subject-specific reasoning.

We investigate this hypothesis across five complementary dimensions: knowledge source ablation (RQ1), statistical context (RQ2), structural property identification (RQ3), perturbation sensitivity (RQ4), and reasoning fidelity (RQ5). Using the ADHD-200~\cite{adhd2012adhd} dataset with six LLMs ranging from small open-weight models to a proprietary model, we find little evidence that LLM predictions are consistently grounded in subject-specific connectivity. The same pattern recurs across the five dimensions: model behavior is strongly affected by surface cues and does not reliably reflect the connectivity provided to them. A manual audit further reveals inconsistent interpretations of the supplied evidence; in one case, the same model interprets the same within-network connectivity count as evidence for Control in one subject but as evidence for ADHD in another.

\noindent\textbf{Our main contributions are as follows:}
\begin{itemize}
\item \textbf{The surface cue hypothesis.} We formalize and test the surface cue hypothesis, turning the question of whether LLMs reason over connectivity or respond to surface cues into a systematic, falsifiable test.

\item \textbf{A five-dimensional probing study.} We design a rigorous study that varies one factor at a time across five dimensions, combining input ablations, structural perturbations, and a manual audit of the models' own justifications, evaluated on six models ranging from small open-weight LLMs to a proprietary model.

\item \textbf{Evidence challenging a common assumption.} Current LLM--brain methods implicitly assume that model outputs reflect the connectivity provided to them. Our results challenge this assumption: predictions are more consistent with reliance on surface cues than on subject-specific connectivity, and this pattern persists across models of different scales and families.
\end{itemize}

\section{Related Work}
\subsection{LLM-Enhanced Brain Graph Analysis}
Recent studies have begun incorporating LLMs into brain graph analysis by coupling LLMs with functional connectomes in two directions, either injecting LLM-derived semantic knowledge into neuroimaging models\cite{XuJia_BrainPrompt_MICCAI2025, dong2026bleg}, or encoding connectivity into LLMs' semantic space\cite{wei2026fmri, hu2026fcn}. BrainPrompt\cite{XuJia_BrainPrompt_MICCAI2025} integrates LLM-encoded ROI-level, subject-level, and disease-level prompts into a Graph Neural Network (GNN), supplying information about region identity, subject characteristics, and disease progression to enhance neurological condition identification. BLEG\cite{dong2026bleg} similarly uses an LLM to generate a textual representation of brain graphs derived from blood-oxygen-level-dependent (BOLD) signal time series parcellated with the Automated Anatomical Labeling (AAL) atlas, and aligns the language-model representation with a GNN for downstream tasks. More recent methodologies approach direct brain-language integration: fMRI-LM\cite{wei2026fmri} converts rs-fMRI derived information, including FC and graph-theoretic measures, into a standardized textual description and utilizes a pretrained LLM to model fMRI tokens alongside text. FCN-LLM\cite{hu2026fcn} aligns multi-scale FC network representations with LLMs through graph-level multi-task instruction tuning, incorporating subject-specific attributes such as demographics, phenotypes, and psychiatric conditions to enable LLM-based interpretation of brain networks. These methods demonstrate that LLMs can contribute useful representations, domain knowledge, or reasoning capabilities to brain graph analysis. Moving beyond representation and alignment, BrainAgent~\cite{li2026language} places the LLM directly in the connectome classification loop through iterative topology-aware description, retrieval, and reasoning. 
What remains unclear is whether LLM outputs on textualized brain graphs are actually grounded in subject-specific connectivity. Existing methods evaluate downstream performance but do not isolate whether predictions arise from connectivity itself or from anatomical priors and other surface-level cues. Although general graph-reasoning studies examine encoding and naming effects~\cite{fatemi2024talk, dai2024revisiting, zhang2024llm4dyg}, this question remains unexplored for brain connectivity graphs. We address this gap through controlled probes that test whether LLM predictions and explanations genuinely reflect the connectivity provided to them.
\begin{figure*}
    \centering
    \includegraphics[width=\textwidth]{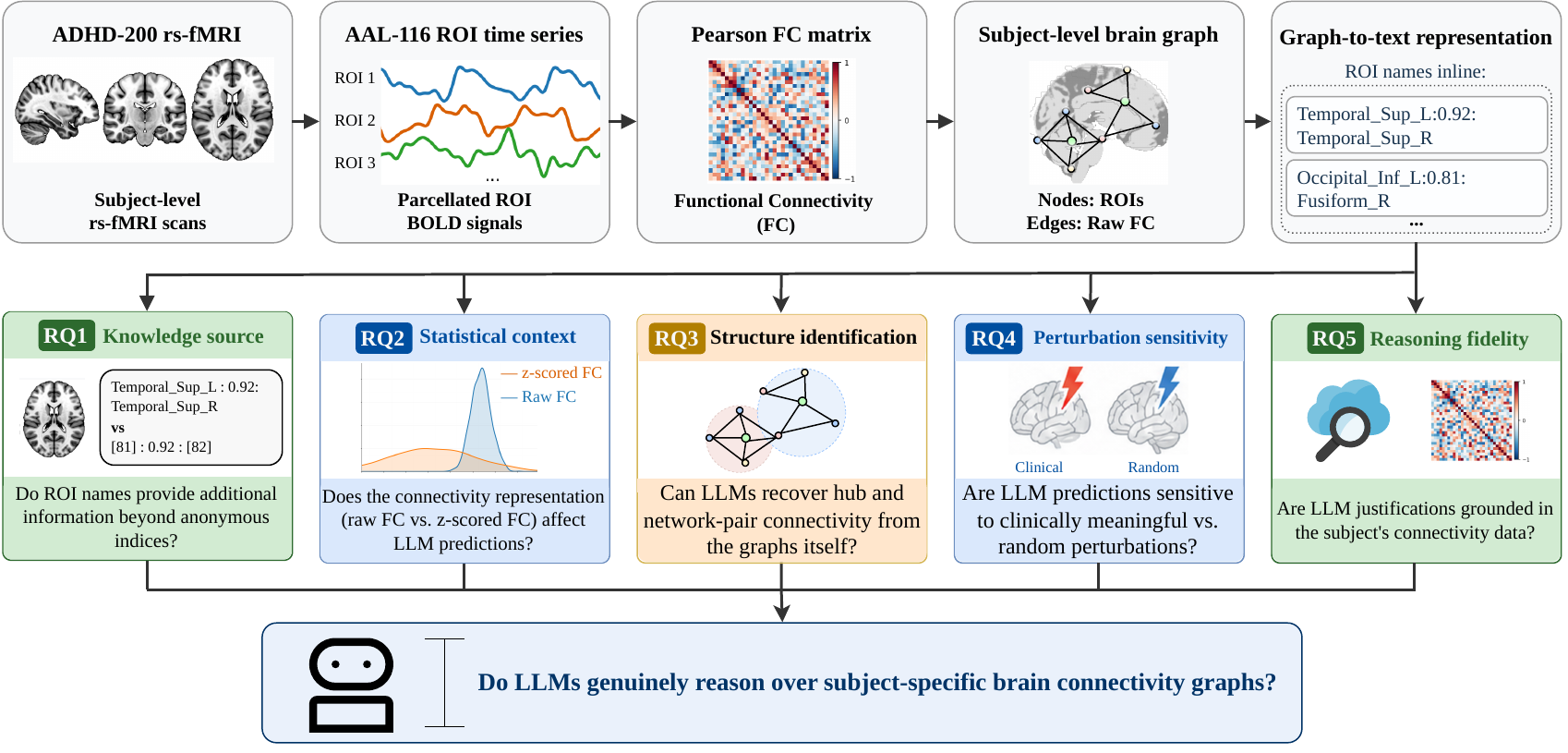}
    \caption{ Overview of the study design for evaluating whether LLMs genuinely reason over subject-specific brain connectivity graphs. The pipeline summarizes brain graph construction and the five complementary probing dimensions: knowledge source, statistical context, structural property identification, perturbation sensitivity, and reasoning fidelity.}
       \label{fig:brianllm_main_diagram}\vspace{-0.5cm}
\end{figure*}

\subsection{Graph Reasoning with LLMs}
The ability of LLMs to reason reliably over graph structure remains an open research problem. NLGraph~\cite{wang2023can} finds that LLMs have limited graph reasoning capabilities, with performance declining as graph tasks become more complex and sensitivity to irrelevant information increases. GraphArena~\cite{tang2025grapharena} extends this to larger real-world graphs, reporting that even leading models struggle as graph size increases and frequently produce hallucinated outputs. Talk Like Graph~\cite{fatemi2024talk} investigates how encoding a graph as text can affect measured performance, with encoder choice alone shifting reasoning performance by 4.8\% to 61.8\%. These representation effects are model-dependent and may not generalize across models~\cite{sclar2024quantifying}. A related challenge concerns numerical information. Numerologic~\cite{schwartz2024numerologic} links LLMs' numerical weakness to the non-intuitive textual representation of numbers, where place value emerges only after reading the full number. Similar sensitivity has been observed in clinical settings, where performance varies across semantically equivalent representations of the same clinical values and shows weaknesses in relational comparison and aggregation over subject-specific information~\cite{nguyen2026robust}.

NLGift~\cite{zhang2024can} further shows that graph instruction-tuned LLMs can generalize across some semantic and numerical patterns but struggle with real-world patterns, raising concerns that their performance may reflect pattern memorization rather than transferable graph reasoning. Beyond task performance, the reliability of LLM explanations presents another challenge. Turpin et al.~\cite{turpin2023language} show that chain-of-thought rationales can be plausible yet misrepresent the computation underlying an answer. Similar concerns have been reported in clinical applications, where generated rationales may not faithfully reflect model decision-making~\cite{afolabi2026faithful}, and predictions can depend on superficial cues and spurious correlations rather than task-relevant evidence~\cite{sakib2025spurious}. These findings are particularly relevant to brain graphs, where textual representations expose both ROI identities and numerical connectivity values. LLMs may therefore rely on semantic cues from ROI labels rather than reasoning over the associated connectivity. This raises a key question: whether apparent reasoning over brain graphs reflects sensitivity to subject-specific connectivity or reliance on surface-level cues.

\subsection{Brain Graph Properties as Biomarkers}
Graph-theoretic analysis of FC has established several network properties as neural correlates of neurodevelopmental and psychiatric conditions~\cite{bullmore2009complex}, with particular relevance to ADHD~\cite{wang2009altered,qian2019large}. These properties describe how the brain's network is organized, both at the level of individual regions and across functional systems. We probe two of them, hub structure and network-pair connectivity, because both have documented ADHD signatures and can be derived directly from a subject's connectivity graph.

Hubs are highly connected regions that facilitate information flow across the brain~\cite{rubinov2010complex,sporns2007identification}. Their organization is altered in ADHD: degree differences recur across multiple studies, particularly in cingulate regions, and have been used for automated classification~\cite{dey2012exploiting,yao2014functional}. At the network level, ADHD is associated with altered connectivity both within and between functional systems~\cite{guo2020shared}, including the default-mode, frontoparietal control, and salience/ventral-attention networks~\cite{cai2018aberrant,menon2011large}, which correspond to the Default, Cont, and SalVentAttn networks of the Yeo-7 taxonomy~\cite{yeo2011organization}. Because both hub structure and network-pair connectivity can be derived from the subject-level graph, we can test whether LLMs identify structural properties that are relevant to ADHD.

\section{Experiments}
We investigate whether LLMs genuinely reason over brain connectivity or rely on surface cues. Fig.~\ref{fig:brianllm_main_diagram} provides an overview of our experimental pipeline, from subject-level rs-fMRI and brain graph construction to the five complementary probing dimensions. We first describe the dataset and graph construction (Section III-A), then the design of our five probes (Section III-B), and finally the language models evaluated (Section III-C).
\begin{table*}
\centering
\caption{Six input conditions for RQ1 and RQ2, all using the same example edge (Temporal\_Sup\_L $\leftrightarrow$ Temporal\_Sup\_R). Columns vary how ROI identity is shown; rows vary whether the value is a raw FC or a cohort $z$-scored FC. In Scn.\,2 and Scn.\,5, ROI identities are given by a separate node-index legend in the prompt rather than in the edge itself (e.g., \texttt{Node 81=Temporal\_Sup\_L, Node 82=Temporal\_Sup\_R, ...}). Scn. stands for scenario.}
\label{tab:rq1_rq2_conditions}

\small
\setlength{\tabcolsep}{9pt}
\renewcommand{\arraystretch}{1.35}

\begin{tabular}{@{}l c c c@{}}
\toprule
&
\textbf{No ROI names} &
\textbf{Names in prompt} &
\textbf{Names inline}
\\
\midrule

\textcolor{fcblue}{\textbf{Raw FC value}}
&
\textbf{Scn.\,1}
&
\textbf{Scn.\,2}
&
\textbf{Scn.\,3}
\\[-1pt]

&
\texttt{[81]:\colorbox{zblue}{0.92}:[82]}
&
\makecell{
\texttt{[81]:\colorbox{zblue}{0.92}:[82]} \\[2pt]
{\scriptsize\color{black}Node 81 = Temporal\_Sup\_L, ...}
}
&
\texttt{Temporal\_Sup\_L:\colorbox{zblue}{0.92}:Temporal\_Sup\_R}
\\

\midrule

\textcolor{tgreen}{\textbf{$z$-scored FC}}
&
\textbf{Scn.\,4}
&
\textbf{Scn.\,5}
&
\textbf{Scn.\,6}
\\[-1pt]

&
\texttt{[81]:\colorbox{zgreen}{1.38}:[82]}&
\makecell{
\texttt{[81]:\colorbox{zgreen}{1.38}:[82]} \\[2pt]
{\scriptsize\color{black}Node 81 = Temporal\_Sup\_L, ...}
}
&
\texttt{Temporal\_Sup\_L:\colorbox{zgreen}{1.38}:Temporal\_Sup\_R}
\\

\bottomrule
\end{tabular}\vspace{-0.4cm}
\end{table*}
\subsection{Dataset and Graph Construction}
\label{sec:dataset}
We use ADHD-200~\cite{adhd2012adhd}, a multi-site, open-access rs-fMRI collection. We use the Athena-pipeline derivatives from the Neuro Bureau's ADHD-200 Preprocessed repository~\cite{bellec2017neuro}, obtained through the Neuroimaging Informatics Tools and Resources Clearinghouse (NITRC). The released data provide preprocessed regional mean time series based on the AAL atlas~\cite{tzourio2002automated}, comprising $N_r=116$ ROIs per subject; the Brown University site is not included in this release because diagnostic labels were unavailable~\cite{bellec2017neuro}. We further exclude 8 subjects whose fMRI files contain zero-valued (empty) columns. The resulting cohort comprises $N_s=768$ subjects across two diagnostic classes: ADHD (280) and Control (488), with the three ADHD subtypes collapsed into the ADHD class.

Each subject is represented as an undirected weighted graph $\mathcal{G} = (\mathcal{V}, \mathcal{E}, W)$. The node set $\mathcal{V} = \{v_1, \dots, v_{N_r}\}$ consists of the $N_r$ AAL ROIs, one node per region. Where multiple runs are available, run-level series are averaged element-wise into a single subject-level series. Let $\mathbf{x}_i \in \mathbb{R}^{T}$ denote the BOLD time series associated with node $v_i$ over $T$ time points. For subject $s$, the weight matrix $W^{(s)} \in \mathbb{R}^{N_r \times N_r}$ assigns to each region pair the raw Pearson FC between their time series~\cite{rubinov2010complex}:
\begin{equation} W^{(s)}_{ij} = \text{corr}(\mathbf{x}_i, \mathbf{x}_j), \end{equation}
giving a dense matrix over all $\binom{116}{2}=6{,}670$ region pairs. Since positive and negative correlations exhibit distinct network properties~\cite{schwarz2011negative}, we set negative FC values to zero. The edge set $\mathcal{E}$ is then obtained by sparsification: an edge $e_{ij}$ connecting nodes $v_i$ and $v_j$ exists if and only if $W^{(s)}_{ij}$ ranks among the top-10\% of that subject's positive values, yielding $|\mathcal{E}|=667$ edges per subject. This edge set is serialized into a textual edge list of the form \texttt{A:value:B} and serves as the graph-to-text representation for all experiments; the specific value encoding and whether region names are shown vary by condition (Section~\ref{subsection:rq1_2_setup}).

\subsection{Probe Design Across RQs}
Our five research questions (RQs) build on one another, but each stands on its own: a later question uses what an earlier one reveals, yet stays meaningful. RQ1 and RQ2 work from the same connectivity but probe different things: whether the model leans on the anatomical names shown alongside those values (RQ1, knowledge source), and whether representing the values as cohort $z$-scored FC rather than raw correlations changes how it uses them (RQ2, statistical context). RQ3 tests whether it can recover real structure, hubs, and network-pair connectivity from the graph. RQ4 then perturbs that structure to see whether the model reacts to clinically meaningful changes or to any change at all. RQ5 steps back and checks the model's stated justifications against each subject's own data.
One caution runs through all five: on an imbalanced cohort, overall accuracy can be misleading, since a model that simply favors the majority class looks accurate without reading the graph at all. We therefore track class-wise behavior alongside overall accuracy throughout. We prompt the LLMs across a range of input conditions and require every response to follow a structured format. We describe the setup for each RQ below.

\subsubsection{\textbf{Knowledge Source and Statistical Context}}
\label{subsection:rq1_2_setup}
We first ask whether region names help an LLM's ADHD/Control prediction, or whether the model uses the connectivity and its values instead. Because anatomical names may cue disorder associations on their own (Section~\ref{sec:intro}), we hold the connectivity values fixed and vary only how much region identity the prompt reveals, across three scenarios. In the first, nodes appear as numeric indices alone, so the model sees connectivity without any anatomical identity. In the second, the region names are listed separately in the prompt while the edges still use indices. In the third, the names are written directly into each edge. Comparing the three isolates the effect of name exposure while the underlying graph stays the same (Table~\ref{tab:rq1_rq2_conditions}).

We then ask whether presenting the connectivity values differently helps the model use them, rather than falling back on the names. A raw correlation value between ROIs is hard to interpret on its own: the same value can be strong for one edge and weak for another, depending on how that edge behaves across the cohort. A cohort-level $z$-score removes this ambiguity by expressing each value relative to its own distribution, making its magnitude directly comparable across edges and subjects. We repeat all three name scenarios with $z$-scored FC in place of raw FC values, giving six input variants in total (Table~\ref{tab:rq1_rq2_conditions}). The edge set is identical across all six; only the displayed value changes.

The $z$-scored FC is computed over the same edge set $\mathcal{E}$ defined in Section~\ref{sec:dataset} (the top-10\% positive edges). For each edge $e$ corresponding to region pair $(i, j)$, let $S_e = \{\, s: W^{(s)}_{ij} > 0 \,\}$ be the set of subjects with positive raw FC on that edge. The cohort mean and standard deviation over $S_e$ are
\[
\mu_e = \frac{1}{|S_e|}\sum_{s \in S_e} W^{(s)}_{ij},
\qquad
\sigma_e = \sqrt{\frac{1}{|S_e|}\sum_{s \in S_e}\left(W^{(s)}_{ij}-\mu_e\right)^2},
\]
and the cohort-relative $z$-score is $
z_{s,e} = \frac{W^{(s)}_{ij}-\mu_e}{\sigma_e + \epsilon}, \qquad \epsilon = 10^{-8},
$ with $\epsilon$ added for numerical stability. Comparing the six variants isolates two effects: name exposure (comparing the three name conditions while holding the value type fixed) and numeric representation (comparing raw FC values versus $z$-scored FC while holding the name condition fixed). We evaluate each variant on binary ADHD/Control classification, reporting overall accuracy and class-wise recall.

\subsubsection{\textbf{Structural Property Identification}}
\label{subsection:rq_3_setup}
Beyond predicting a label from connectivity values, we examine whether an LLM can recover the graph's structural organization directly from the edge list, rather than relying on prior associations with ADHD. We probe two complementary properties: hubs, which capture the connectivity prominence of individual regions, and network-pair connectivity, which captures how many retained edges connect functional systems within and between networks. Both are established markers of network organization and are reported to differ in ADHD~\cite{cai2018aberrant, wang2009altered, qian2019large}, making them natural probes of whether the model reads subject-specific structure rather than generic priors.

For each property, the model receives the subject's unweighted edge list $\mathcal{E}$ with ROI names, together with a one-shot example following~\cite{tang2025grapharena}, and is asked to identify the hub or network-pair connectivity. Each prediction is scored by its overlap with an independently computed ground truth. We first validate predictions against the task's admissible label set: outputs naming invalid regions or network-pairs are classified as \textit{hallucinatory}, and those that cannot be parsed into a structural prediction as \textit{missing}. Valid predictions are graded by overlap as \textit{correct} ($\geq 0.6$), \textit{partial} ($0.4$--$0.59$), or \textit{suboptimal} (below $0.4$).

\paragraph{Hub Identification}
Hubs are the most heavily connected regions in a graph, acting as central relay points for information flow; their disruption is a documented correlate of ADHD~\cite{sporns2007identification, yao2014functional}. We define hubs by node degree. The degree of node $v_i$ is the number of edges incident to it in $\mathcal{E}$, $k_i = |\{j : e_{ij} \in \mathcal{E}\}|$. We rank all nodes by degree and take the top 20 as the ground-truth hub set:
\begin{equation}
\mathcal{H} = \operatorname{top\text{-}20}_{v_i \in \mathcal{V}} \; k_i,
\end{equation}
yielding $|\mathcal{H}| = 20$ hubs per subject. The input is provided to the LLM as an unweighted edge list with ROI names
(e.g., \texttt{Temporal\_Sup\_L to Temporal\_Sup\_R}). The LLM is instructed to return the 20 highest-degree regions from the subject's edge list, and its prediction is validated against $N_r$ region names, then compared against $\mathcal{H}$.

\paragraph{Network-Pair Connectivity}
Beyond individual regions, ADHD is associated with altered connectivity between functional systems. We use the canonical Yeo-7 networks as a fixed partition and test whether the model can read off how connectivity is distributed across them.

Most of the 116 AAL ROIs are assigned to one of the seven Yeo functional networks~\cite{yeo2011organization} based on the Yeo network containing the largest number of voxels within each ROI, after resampling the Yeo-7 atlas into AAL space. ROIs with no Yeo-7 overlap are retained as ``Unassigned,'' yielding $M=8$ network labels in total. We denote the seven Yeo networks as Visual, SomMot, DorsAttn, SalVentAttn, Limbic, Cont, and Default throughout the analysis. We label every edge with the networks of its two endpoints and give the model the edge list in this labeled form, e.g., \texttt{Temporal\_Sup\_L (SomMot) to Temporal\_Sup\_R (SomMot)}. The model's task is to aggregate these labeled edges and report which network-pairs are most connected.

Let $g(v_i)$ be the Yeo network label of ROI $v_i$, one of the $M=8$ networks. For any two networks $m$ and $n$ (for example, default-mode and somatomotor), the number of edges connecting them is
\begin{equation}
c_{mn} =
\sum_{(i,j)\in\mathcal{E}}
\mathbf{1}\left[
\{g(v_i),g(v_j)\}=\{m,n\}
\right],
\end{equation}
where $\mathbf{1}[\cdot]$ equals 1 when its condition holds and 0 otherwise. Taking $m=n$ gives the 8 within-network counts and $m \neq n$ the $\binom{8}{2}=28$ between-network counts, 36 in total. Ranking these in descending order gives the ground-truth ordering, and the model is asked to return the top-10 network pairs, scored by overlap with the ground-truth top-10.

\subsubsection{\textbf{Perturbation Design}}
\label{subsection:pertubation_design}
To assess whether an LLM is sensitive to the clinically meaningful parts of a graph, we perturb each subject's hub list and observe how its prediction responds. Starting from the ground-truth top-20 hub set $\mathcal{H}$, we replace $K$ of its members with non-hub regions (ranked 21st-116th by degree), keeping the list at 20. We vary the magnitude with $K \in \{5, 10\}$, expecting larger $K$ to disrupt predictions more.

The two conditions differ in which hubs are replaced. In the \textbf{clinical} condition, replacements come only from the three networks most implicated in ADHD (default-mode, frontoparietal control, and salience/ventral-attention~\cite{cai2018aberrant,menon2011large}, corresponding to \textit{Default}, \textit{Cont}, and \textit{SalVentAttn} in the Yeo-7 taxonomy~\cite{yeo2011organization}), the regions a clinically grounded reasoner would rely on. In the \textbf{random} condition, replacements are drawn uniformly at random. Since the conditions differ only in this clinical specificity, a model that truly reads clinically relevant structure should be disturbed more by the clinical perturbation than the random one.

We run every perturbation under three prompt formulations that add progressively more context: (i) the hub list alone, with each hub's degree and cohort $z$-scored FC (H); (ii) the same list plus prompt-provided prior knowledge of ADHD-implicated networks (H+P); and (iii) the same list plus each hub's Yeo-7 network label and that prior knowledge (H+P+N). These let us test whether any sensitivity comes from the graph itself or only appears once the LLM is told which regions are clinically relevant.

To measure a perturbation's effect, we compare the LLM predictions before and after it with two complementary measures: the \textbf{flip rate}, which captures the \emph{magnitude} of change, and \textbf{McNemar's test}, which captures its \emph{direction}.
\paragraph{Flip Rate}
The flip rate captures the \textit{magnitude} of change between two prediction states. We compare three pairs:  baseline (unperturbed) vs.\ clinical, baseline vs.\ random, and random vs.\ clinical. For each pair, it is the proportion of scorable subjects whose predicted label differs between the two states. Let $N_{\text{flipped}}$ be the number of such subjects, and $N_{\text{scorable}}$ be the number of valid predictions, excluding subjects that are missing or incorrectly formatted: $\text{Flip Rate} = \frac{N_{\text{flipped}}}{N_{\text{scorable}}}.$

\paragraph{McNemar's Test}
The flip rate counts how many predictions change, but not in which direction. McNemar's test~\cite{mcnemar1947note} captures direction: for a pair of conditions, let $b$ be the subjects correct before the perturbation but wrong after, and $c$ the reverse (unchanged subjects are ignored). If the perturbation had no systematic effect, $b$ and $c$ would be roughly equal. The exact two-sided test~\cite{dietterich1998approximate} scores how unlikely the observed split is:
\begin{equation}
p = 2\sum_{i=0}^{\min(b,c)} \binom{b+c}{i}\left(\frac{1}{2}\right)^{b+c}.
\end{equation}
Here, $\binom{b+c}{i}$ means binomial coefficient. A small $p$ means the flips lean one way; $b > c$ means the perturbation degrades accuracy more than it helps. Together, the two measures test our hypothesis: a model that reads clinically relevant structure should be disturbed more, and more systematically, by the clinical perturbation than the random one.

\subsubsection{\textbf{Reasoning Fidelity}}
\label{subsection:rq5_reasioning}
Correct prediction alone does not demonstrate that a model has reasoned correctly over the graph. A model can arrive at the right label for the wrong reasons; for instance, chain-of-thought explanations may misrepresent the actual factors influencing a prediction~\cite{turpin2023language,afolabi2026faithful}.

Consequently, we go beyond aggregate accuracy and manually audit reasoning fidelity at the subject level. Along with each prediction, the model produces a short justification explaining it. Drawing on the principle of explanation faithfulness discussed by Jacovi and Goldberg~\cite{jacovi2020towards}, we examine whether each justification is consistent with the subject-specific connectivity evidence rather than relying on generic statements. These justifications come from a classification task in which the model receives each subject's ground-truth within- and between-network connectivity counts (the aggregated $c_{mn}$ values from Section~\ref{subsection:rq_3_setup}) and is asked to predict ADHD or Control and explain its choice. For each model (Llama, Luna, and DeepSeek-V4), we draw a stratified sample of 10 subjects (5 ADHD, 5 Control) and verify each distinct claim in the justification against these network-pair counts.

\subsection{Language Models}
All experiments are run on post-trained models spanning different families and parameter scales: Llama-3.1-8B-Instruct (Llama)\cite{grattafiori2024llama}, Qwen2.5-14B-Instruct (Qwen2.5)\cite{hui2024qwen2}, Qwen3-14B (Qwen3)\cite{yang2025qwen3}, and DeepSeek-R1-Distill-Qwen-14B (R1-Distill)\cite{guo2025deepseek}. Open-weight models are loaded locally using 4-bit NF4 quantization with double quantization\cite{dettmers2023qlora}. Additionally, we evaluate two API accessed models, DeepSeek-V4-Flash (DeepSeek-V4)\cite{xu2026deepseek}, a 284B-parameter mixture-of-experts model, and GPT-5.6-Luna (Luna)\cite{openai2026gpt56luna}, a closed-weight proprietary model. Both models are queried through their respective provider endpoints and are evaluated without local quantization. Local models used greedy decoding, whereas API-accessed models used provider-default sampling settings. Unparseable responses are labeled \textit{missing} and excluded from $N_{\text{scorable}}$, rather than imputed to either diagnostic class.

\section{Result Analysis}
\subsection{Knowledge Source and Statistical Context (RQ1–RQ2)}
\label{subsection:rq_1_2_result}
Table~\ref{tab:rq12_accuracy} reports overall accuracy (A) and class-wise recall for ADHD (D) and Control(C) across the six input formats and six models. Because the cohort is imbalanced (488 of 768 subjects are Control), predicting Control for every subject yields 63.54\% accuracy; we treat this majority-class baseline as the reference for all results.

\begin{table}[t]
    \centering

    \caption{Classification performance (\%) across six
    input-format scenarios and six LLMs.}

    \label{tab:rq12_accuracy}

    \scriptsize
    \setlength{\tabcolsep}{3pt}
    \renewcommand{\arraystretch}{1.15}

    \begin{tabular}{@{}
        l
        >{\centering\arraybackslash}p{0.105\columnwidth}
        >{\centering\arraybackslash}p{0.135\columnwidth}
        >{\centering\arraybackslash}p{0.125\columnwidth}
        >{\centering\arraybackslash}p{0.115\columnwidth}
        >{\centering\arraybackslash}p{0.135\columnwidth}
        >{\centering\arraybackslash}p{0.125\columnwidth}
        @{}}

        \toprule
        \textbf{Scenario}
        & \textbf{R1- Distill}
        & \textbf{DeepSeek- V4}
        & \textbf{Luna}
        & \textbf{Llama}
        & \textbf{Qwen2.5}
        & \textbf{Qwen3} \\
        \midrule

        \multirow{3}{*}{Scn.~1}
        & A: 51.20
        & A: 52.81
        & \textbf{A: 55.41}
        & \textbf{A: 59.57}
        & A: 63.28
        & \textbf{A: 52.93} \\

        & D: 51.09
        & D: 33.57
        & D: 36.07
        & D: 19.05
        & D: 0.37
        & D: 43.73 \\

        & C: 51.27
        & C: 63.92
        & C: 66.53
        & C: 82.67
        & C: 99.79
        & C: 58.20 \\

        \addlinespace[1pt]
        \cmidrule(lr){1-7}
        \addlinespace[1pt]

        \multirow{3}{*}{Scn.~2}
        & \textbf{A: 53.15}
        & \textbf{A: 57.81}
        & A: 45.89
        & A: 58.20
        & \textbf{A: 63.71}
        & A: 41.05 \\

        & D: 45.56
        & D: 18.21
        & D: 53.21
        & D: 13.45
        & D: 1.10
        & D: 80.43 \\

        & C: 57.44
        & C: 80.53
        & C: 41.68
        & C: 83.98
        & C: 99.17
        & C: 18.60 \\

        \addlinespace[1pt]
        \cmidrule(lr){1-7}
        \addlinespace[1pt]

        \multirow{3}{*}{Scn.~3}
        & A: 50.69
        & A: 55.60
        & A: 54.04
        & A: 50.68
        & A: 63.27
        & A: 50.40 \\

        & D: 28.52
        & D: 26.07
        & D: 33.57
        & D: 52.59
        & D: 0.00
        & D: 56.04 \\

        & C: 63.40
        & C: 72.54
        & C: 65.78
        & C: 49.57
        & C: 100.00
        & C: 47.12 \\

        \addlinespace[2pt]
        \midrule[1.5pt]
        \addlinespace[2pt]

        \multirow{3}{*}{Scn.~4}
        & A: 44.11
        & A: 40.42
        & A: 50.00
        & A: 36.86
        & A: 60.44
        & A: 40.13 \\

        & D: 78.91
        & D: 79.29
        & D: 42.31
        & D: 92.83
        & D: 7.53
        & D: 83.46 \\

        & C: 24.17
        & C: 18.07
        & C: 54.50
        & C: 4.73
        & C: 90.76
        & C: 15.22 \\

        \addlinespace[1pt]
        \cmidrule(lr){1-7}
        \addlinespace[1pt]

        \multirow{3}{*}{Scn.~5}
        & A: 44.27
        & A: 43.94
        & A: 38.22
        & A: 47.25
        & A: 61.33
        & A: 38.26 \\

        & D: 84.50
        & D: 73.84
        & D: 94.51
        & D: 57.41
        & D: 6.79
        & D: 92.31 \\

        & C: 20.68
        & C: 26.84
        & C: 6.07
        & C: 41.51
        & C: 92.62
        & C: 7.97 \\

        \addlinespace[1pt]
        \cmidrule(lr){1-7}
        \addlinespace[1pt]

        \multirow{3}{*}{Scn.~6}
        & A: 39.26
        & A: 42.19
        & A: 37.34
        & A: 42.97
        & A: 62.22
        & A: 41.01 \\

        & D: 74.01
        & D: 73.21
        & D: 96.77
        & D: 67.87
        & D: 2.88
        & D: 78.77 \\

        & C: 19.08
        & C: 24.39
        & C: 3.29
        & C: 28.72
        & C: 96.10
        & C: 18.84 \\

        \bottomrule

    \end{tabular}

    \vspace{2pt}

    \begin{minipage}{\columnwidth}
        \scriptsize

        A denotes overall accuracy, while D and C denote class-specific
        accuracies for ADHD and Control subjects, respectively.
        Bold values indicate the highest overall accuracy for each model.

    \end{minipage}
    \vspace{-6mm}
\end{table}

\textbf{ROI-name exposure has an inconsistent effect}
Across the three raw FC scenarios (Table~\ref{tab:rq1_rq2_conditions}, Scn.\ 1-3), which expose ROI names at increasing levels, no names (Scn.\ 1), names listed separately in the prompt (Scn.\ 2), and names inline in each edge (Scn.\ 3), overall accuracy shifts by as little as 0.44 points for Qwen2.5 but by 8.89 and 11.88 points for Llama and Qwen3. The direction is inconsistent as well: more name exposure raises ADHD recall for some models while lowering it for others (e.g., Qwen3 fluctuates $43.73 \to 80.43 \to 56.04$ across the three, whereas R1-Distill falls $51.09 \to 28.52$). Anatomical naming therefore provides no consistent advantage: the effect varies across models, with no shared pattern that would indicate a common anatomical prior.

\textbf{$z$-scoring shifts class bias without improving discrimination}
Each raw FC scenario has a corresponding $z$-scored condition with the same ROI-name exposure (Scn.\ 1 vs.\ 4, 2 vs.\ 5, and 3 vs.\ 6), allowing us to examine how predictions change when raw connectivity values are replaced by cohort-relative $z$-scores. Across all three pairs, $z$-scoring raises ADHD recall and lowers Control recall for every model (Table~\ref{tab:rq12_accuracy}). The shift is often large: Llama moves from 19.05\% to 92.83\% ADHD recall, while its Control recall drops from 82.67\% to 4.73\% (Scn.\ 1 vs.\ 4). However, overall accuracy drops under $z$-scoring for every model, while balanced accuracy remains near chance. Thus, $z$-scoring systematically shifts predictions toward the minority (ADHD) class without improving discrimination.

\textbf{High accuracy can mask majority-class collapse}
Qwen2.5 illustrates how overall accuracy can be misleading under class imbalance. In Scn.\ 2, it achieves the highest overall accuracy (63.71\%), yet recalls 99.17\% of Control subjects and only 1.10\% of ADHD subjects. In Scn.\ 3, this imbalance becomes complete: the model predicts every subject as Control, yielding 100\% Control recall and 0\% ADHD recall while still achieving 63.27\% accuracy. These results show that its apparent accuracy largely reflects the majority-class distribution rather than discrimination between ADHD and Control subjects. We therefore report class-wise recall alongside overall accuracy throughout.

\textbf{The pattern persists across model scales}
The preceding effects are not confined to the smaller open-weight models. The larger models show the same qualitative pattern: ROI-name exposure does not yield consistent improvement, while $z$-scoring shifts class preference without improving discrimination. Such as, DeepSeek-V4's overall accuracy drops from 52.81\% to 40.42\% between Scn.\ 1 and Scn.\ 4. Thus, these patterns persist across models of different scales and families.

\begin{table}[t]
\centering
\caption{Structural identification: number of subjects per outcome category
($N_s=768$). The largest count in each row is
 shown in bold. Subopt. stands for Suboptimal and Halluc. stands for Hallucinatory.}
\label{tab:rq3_identification}
\footnotesize
\setlength{\tabcolsep}{5pt}
\renewcommand{\arraystretch}{1.2}
\begin{tabular}{@{}lrrrrr@{}}
\toprule
\textbf{Model} & \textbf{Correct} & \textbf{Partial} & \textbf{Subopt.} & \textbf{Halluc.} & \textbf{Missing} \\
\midrule
\multicolumn{6}{@{}l}{\emph{Hub identification}} \\

R1-Distill
& 9 & 85 & \textbf{367} & 177 & 130 \\

DeepSeek-V4
& 307 & \textbf{391} & 44 & 26 & 0 \\

Luna
& \textbf{593} & 169 & 2 & 1 & 3 \\

Llama 
& 0 & 26 & 222 & \textbf{520} & 0 \\

Qwen2.5 
& 117 & 298 & \textbf{348} & 5 & 0 \\

Qwen3 
& 30 & 160 & \textbf{536} & 42 & 0 \\

\midrule
\multicolumn{6}{@{}l}{\emph{Network-Pair Connectivity}} \\

R1-Distill 
& 15 & 51 & \textbf{515} & 3 & 184 \\

DeepSeek-V4 
& \textbf{666} & 89 & 1 & 12 & 0 \\

Luna
& \textbf{751} & 2 & 0 & 0 & 15 \\

Llama & 90 & \textbf{342} & 335 & 1 & 0 \\

Qwen2.5 
& 70 & \textbf{476} & 222 & 0 & 0 \\

Qwen3 
& \textbf{417} & 312 & 38 & 1 & 0 \\

\bottomrule
\end{tabular}\vspace{-0.3cm}
\end{table}

\begin{figure*}
\centering
\includegraphics[width=\textwidth,height=4.2cm]{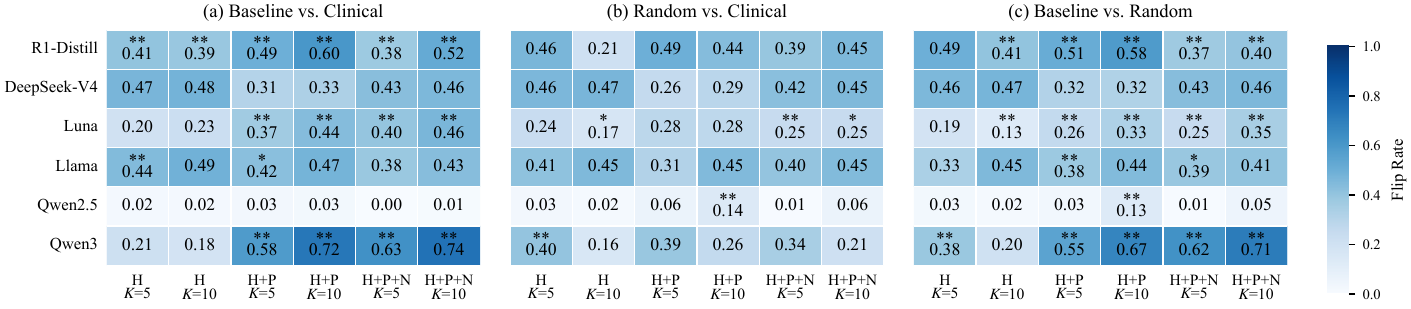}
\caption{Effect of hub perturbation across models, prompt formats, and perturbation magnitudes. Each cell reports the \textbf{flip rate}, defined as the proportion of scorable subjects whose predicted label changes between the two conditions, with asterisks indicating a significant directional difference under the exact McNemar test ($^{*}p<0.05$, $^{**}p<0.01$). Panels compare the unperturbed baseline with clinical (a) and random (c) perturbations, and random with clinical perturbations (b). Prompt formats: H = hub list only; H+P = hub list with ADHD-related prior knowledge; H+P+N = H+P with each hub's Yeo-7 network label. $K$ denotes the number of perturbed hubs.}
\label{fig:flip_heatmaps}\vspace{-1.0cm}
\end{figure*}

\begin{table*}
\centering
\caption{Representative cases of reasoning inconsistency across Luna, Llama, and DeepSeek-V4. Each model is given the subject's ground-truth within- and between-network connectivity counts for classification. GT stands for ground truth; Pred. stands for predicted class.}
\label{tab:reasoning_fidelity}

\small
\renewcommand{\arraystretch}{1.18}
\setlength{\tabcolsep}{4.5pt}

\begin{tabular}{
@{}
>{\raggedright\arraybackslash}p{2.2cm}
>{\centering\arraybackslash}p{1cm}
>{\raggedright\arraybackslash}p{1.8cm}
>{\centering\arraybackslash}p{1.0cm}
>{\centering\arraybackslash}p{1.0cm}
>{\raggedright\arraybackslash}p{2.5cm}
>{\raggedright\arraybackslash}p{6.1cm}
@{}
}
\toprule
\textbf{Case} &
\textbf{Subject} &
\textbf{Model} &
\textbf{GT} &
\textbf{Pred.} &
\textbf{Relevant Connectivity Evidence Counts} &
\textbf{Justification Excerpt} \\
\midrule

\textbf{a:} Same subject, opposite predictions
&
1743472
&
Luna
&
ADHD
&
ADHD
&
DorsAttn (1), SalVentAttn (12), Cont (19) within-network
&
\textit{markedly sparse within-network connectivity in DorsAttn, SalVentAttn, and Control networks} ... \textit{weak attention/control-network cohesion} ... more consistent with an ADHD-like network organization.
\\[3pt]

&
1743472
&
Llama
&
ADHD
&
Control
&
Visual within-network (89)
&
\textit{high within-network connectivity in the Visual network (89)} ... suggests a more typical brain network pattern, indicating a higher likelihood of being classified as Control.
\\

\midrule

\textbf{b:} Same connectivity value, opposite interpretations
&
10129
&
DeepSeek-V4
&
ADHD
&
Control
&
DorsAttn within-network (1)
&
\textit{dorsal attention network is nearly disconnected}; this balanced, integrated profile is more typical of a healthy Control.
\\[3pt]

&
2574674
&
DeepSeek-V4
&
Control
&
ADHD
&
DorsAttn within-network (1)
&
\textit{extremely low within-network connectivity in DorsAttn (1)} ... suggesting reduced segregation and poor attentional network integrity typical of ADHD.
\\

\midrule

\textbf{c:} Similar explanations, different ground truth
&
1743472
&
Luna
&
ADHD
&
ADHD
&
DorsAttn (1), SalVentAttn (12), Cont (19) within-network
&
\textit{markedly sparse within-network connectivity in DorsAttn, SalVentAttn, and Control networks} ... more consistent with an ADHD-like network organization.
\\[3pt]

&
15013
&
Luna
&
Control
&
ADHD
&
DorsAttn (1), SalVentAttn (15), Cont (22) within-network
&
\textit{markedly reduced within-network connectivity in DorsAttn, SalVentAttn, and Control networks} ... more consistent with ADHD than Control.
\\

\midrule

\textbf{d:} Correct prediction, inconsistent evidence interpretation
&
1139030
&
Llama
&
Control
&
Control
&
Default-Limbic between-network (90; high in cohort)
&
\textit{between-network connectivity degree counts are relatively low}, suggesting weak connections between networks ... characteristic of Control subjects.
\\

\bottomrule
\end{tabular}\vspace{-0.2cm}
\end{table*}

\subsection{Structural Property Identification (RQ3)}
\label{subsection:rq3_result}Structural identification performance depends heavily on both the model and the type of structure (Table~\ref{tab:rq3_identification}). Hub identification separates the models sharply. Luna recovers the correct hub set for 593 of 768 subjects (77.2\%) and DeepSeek-V4 for 307 (40.0\%), but the remaining models fall away quickly: Qwen2.5 reaches only 15.2\%, and Qwen3, R1-Distill, and Llama manage 3.9\%, 1.2\%, and 0.0\%. The weaker models also fail in different ways. Qwen3 and R1-Distill land mostly in the \textbf{suboptimal} band (69.8\% and 47.8\% of subjects).

Finding the most connected network-pairs is easier for five of the six models. Luna and DeepSeek-V4 improve from 77.2\% and 40.0\% correct hub identification to 97.8\% and 86.7\% correct network-pair identification, respectively, while Qwen3 shows the largest gain, from 3.9\% to 54.3\%. R1-Distill and Llama also improve on the network-pair task. Qwen2.5 is the only exception, decreasing from 15.2\% to 9.1\%. Hallucination shows a different pattern: hub predictions contain invalid regions for 23.0\% of R1-Distill subjects and 67.7\% of Llama subjects, whereas hallucination remains below 2\% for every model on the network-pair task. This difference may partly reflect the smaller output space, with 36 admissible network-pair labels compared with 116 candidate regions for hub identification.
Overall, structural identification varies substantially across models and structure types, with Luna and DeepSeek-V4 recovering the target structures far more reliably than the other models.

\subsection{Perturbation Sensitivity (RQ4)}
\label{subsection:rq4_result}
Figure~\ref{fig:flip_heatmaps} reports flip rates and McNemar tests across all models, prompt formats, and perturbation magnitudes.

\textbf{Perturbation sensitivity varies substantially across models.}
Figures~\ref{fig:flip_heatmaps}(a) and (c) compare the unperturbed baseline with clinical and random perturbations, respectively. The models respond very differently to hub perturbations. Qwen3 is the most sensitive, with flip rates reaching 0.74, whereas Qwen2.5 remains largely unchanged, with flip rates between 0.00 and 0.14. R1-Distill and Llama show significant changes in 11 and 6 of the 12 baseline comparisons, respectively, while DeepSeek-V4 shows no significant baseline-to-perturbation comparison despite moderate flip rates. Thus, perturbing the hub set can substantially alter predictions, but the degree of sensitivity is highly model-dependent.

\textbf{Clinical perturbations are not consistently more disruptive than random perturbations.} If model predictions depend preferentially on clinically implicated hubs, replacing those hubs should produce systematically different effects from replacing randomly selected hubs. Figure~\ref{fig:flip_heatmaps}(b)  provides little evidence for such selectivity. Only 5 of the 36 matched clinical-versus-random comparisons show a significant directional difference. Thus, although some models are sensitive to hub perturbation, this sensitivity does not consistently distinguish clinically implicated hubs from random ones.

\textbf{Sensitivity does not consistently increase with perturbation magnitude.}
Across all three panels of Fig.~\ref{fig:flip_heatmaps}, increasing the number of perturbed hubs from $K=5$ to $K=10$ does not systematically increase flip rates or strengthen the difference between clinical and random perturbations. Across models and prompt formats, some conditions show larger effects at $K=10$, whereas others remain similar or decrease. Thus, greater perturbation does not produce a consistent increase in model sensitivity.
Together, these results show that models can be sensitive to hub perturbations, but this sensitivity is neither consistently selective to clinically implicated hubs nor stronger under larger perturbations.

\subsection{Reasoning Fidelity (RQ5)}
\label{subsection:rq5_result}

For RQ5, each model receives a subject's ground-truth within- and between-network connectivity counts and produces a predicted label together with a short justification (Section~\ref{subsection:rq5_reasioning}). We examine whether the evidence used in each justification is consistent with the supplied connectivity values. Table~\ref{tab:reasoning_fidelity} presents representative cases from this manual audit. On the shared balanced sample of 10 subjects (5 ADHD and 5 Control), Luna predicts ADHD for every subject, whereas Llama predicts Control for every subject; consequently, both achieve 50\% accuracy despite collapsing to a single class. DeepSeek-V4 achieves 30\% accuracy on its sampled subjects. Beyond prediction accuracy, the audit reveals several inconsistencies in how the models use and interpret the supplied connectivity.

\textbf{Same subject, opposite predictions.}
For subject 1743472, Luna and Llama receive the same connectivity profile but reach opposite predictions. Luna predicts ADHD while emphasizing sparse connectivity in the attention and Control networks, whereas Llama predicts Control while emphasizing the high Visual within-network count of 89. Thus, the two models draw different conclusions from the same subject-specific connectivity profile (Table~\ref{tab:reasoning_fidelity}a).

\textbf{Same connectivity value, opposite interpretations.}
The inconsistency also appears within a single model. DeepSeek-V4 receives the same DorsAttn within-network count of 1 for subjects 10129 and 2574674, yet associates it with a Control prediction for the former and an ADHD prediction for the latter. The corresponding explanations characterize this evidence differently, linking it to a ``balanced, integrated profile'' in one case and to ``extremely low'' connectivity typical of ADHD in the other (Table~\ref{tab:reasoning_fidelity}b).

\textbf{Similar explanations across different subjects.}
Luna gives similar explanations for subjects 1743472 and 15013 despite their different ground-truth labels. In both cases, it emphasizes sparse within-network connectivity in DorsAttn, SalVentAttn, and Cont as evidence for ADHD, and consequently predicts ADHD for both subjects (Table~\ref{tab:reasoning_fidelity}c). This suggests that the generated explanation is not consistently adapted to the subject-specific connectivity profile.

\textbf{Correct prediction with inconsistent evidence interpretation.}
Reasoning inconsistency can occur even when the final prediction is correct. Llama correctly classifies subject 1139030 as Control, but describes its Default-Limbic between-network count of 90 as ``relatively low,'' even though this value is high relative to the cohort (Table~\ref{tab:reasoning_fidelity}d). Thus, prediction correctness alone does not ensure that the accompanying justification accurately reflects the supplied connectivity.

Taken together, these cases show inconsistent interpretations of connectivity evidence, providing little evidence that justifications are reliably grounded in subject-specific connectivity.
\section{Conclusions}
In this study, we examine whether LLMs reason over subject-specific brain connectivity graphs or respond to surface-level cues such as region names, numeric formatting, and prompt framing. Across six models on ADHD-200, balanced accuracy remained near chance: representing connectivity as $z$-scores shifted class recall without improving discrimination, while exposing region names had inconsistent effects. Models were sensitive to hub perturbations, but clinically targeted changes were not consistently more disruptive than random ones. Manual audits further revealed inconsistent interpretations of connectivity evidence and similar explanations across subjects with different ground-truth labels. Even models that reliably identified hubs and network-pair connectivity did not translate this ability into grounded diagnostic reasoning.

Our study is limited to a single dataset, one atlas, and a small manual audit; future work should expand the audit, evaluate additional models, and consider other disorders and brain representations. Overall, accuracy alone is insufficient to establish that an LLM reasons over brain connectivity. Class-wise evaluation, controlled perturbations, and justification auditing together suggest that model predictions are more consistent with reliance on surface cues than with robust use of subject-specific connectivity.
\section*{Acknowledgment}
Generative AI tools were used for proofreading, sentence and figure enhancement, and grammatical correction.

\bibliographystyle{IEEEtran}
\bibliography{references}

@article{guo2023gpt4graph,
  title={Gpt4graph: Can large language models understand graph structured data? an empirical evaluation and benchmarking},
  author={Guo, Jiayan and Du, Lun and Liu, Hengyu and Zhou, Mengyu and He, Xinyi and Han, Shi},
  journal={arXiv preprint arXiv:2305.15066},
  year={2023}
}

@inproceedings{tang2025grapharena,
  title={Grapharena: Evaluating and exploring large language models on graph computation},
  author={Tang, Jianheng and Zhang, Qifan and Li, Yuhan and Chen, Nuo and Li, Jia},
  booktitle={International Conference on Learning Representations},
  volume={2025},
  pages={48118--48145},
  year={2025}
}

@inproceedings{fatemi2024talk,
  title={Talk like a graph: Encoding graphs for large language models},
  author={Fatemi, Bahare and Halcrow, Jonathan and Perozzi, Bryan},
  booktitle={International conference on learning representations},
  volume={2024},
  pages={43909--43934},
  year={2024}
}

@article{farahani2019application,
  title={Application of graph theory for identifying connectivity patterns in human brain networks: a systematic review},
  author={Farahani, Farzad V and Karwowski, Waldemar and Lighthall, Nichole R},
  journal={frontiers in Neuroscience},
  volume={13},
  pages={585},
  year={2019},
  publisher={Frontiers Media SA}
}

@article{wang2009altered,
  title={Altered small-world brain functional networks in children with attention-deficit/hyperactivity disorder},
  author={Wang, Liang and Zhu, Chaozhe and He, Yong and Zang, Yufeng and Cao, QingJiu and Zhang, Han and Zhong, Qiuhai and Wang, Yufeng},
  journal={Human brain mapping},
  volume={30},
  number={2},
  pages={638--649},
  year={2009},
  publisher={Wiley Online Library}
}

@article{bullmore2009complex,
  title={Complex brain networks: graph theoretical analysis of structural and functional systems},
  author={Bullmore, Ed and Sporns, Olaf},
  journal={Nature reviews neuroscience},
  volume={10},
  number={3},
  pages={186--198},
  year={2009},
  publisher={Nature Publishing Group UK London}
}

@inproceedings{wei2026fmri,
  title={fmri-lm: Towards a universal foundation model for language-aligned fmri understanding},
  author={Wei, Yuxiang and Zhang, Yanteng and Xiao, Xi and Qian, Chengxuan and Wang, Tianyang and Calhoun, Vince D},
  booktitle={Proceedings of the IEEE/CVF Conference on Computer Vision and Pattern Recognition},
  pages={6931--6940},
  year={2026}
}

@InProceedings{XuJia_BrainPrompt_MICCAI2025,
        author = { Xu, Jiaxing AND He, Kai AND Tang, Yue AND Li, Wei AND Lan, Mengcheng AND Dong, Xia AND Ke, Yiping AND Feng, Mengling},
        title = { { BrainPrompt: Multi-Level Brain Prompt Enhancement for Neurological Condition Identification } },
        booktitle = {proceedings of Medical Image Computing and Computer Assisted Intervention -- MICCAI 2025},
        year = {2025},
        publisher = {Springer Nature Switzerland},
        volume = {LNCS 15971},
        month = {September},
        page = {172 -- 182}
}

@article{dong2026bleg,
  title={BLEG: LLM Functions as Powerful fMRI Graph-Enhancer for Brain Network Analysis},
  author={Dong, Rui and Wang, Zitong and Li, Jiaxing and Zheng, Weihuang and Kong, Youyong},
  journal={arXiv preprint arXiv:2604.07361},
  year={2026}
}

@article{tzourio2002automated,
  title={Automated anatomical labeling of activations in SPM using a macroscopic anatomical parcellation of the MNI MRI single-subject brain},
  author={Tzourio-Mazoyer, Nathalie and Landeau, Brigitte and Papathanassiou, Dimitri and Crivello, Fabrice and Etard, Octave and Delcroix, Nicolas and Mazoyer, Bernard and Joliot, Marc},
  journal={Neuroimage},
  volume={15},
  number={1},
  pages={273--289},
  year={2002},
  publisher={Elsevier}
}

@article{qian2019large,
  title={Large-scale brain functional network topology disruptions underlie symptom heterogeneity in children with attention-deficit/hyperactivity disorder},
  author={Qian, Xing and Castellanos, Francisco Xavier and Uddin, Lucina Q and Loo, Beatrice Rui Yi and Liu, Siwei and Koh, Hui Li and Poh, Xue Wei Wendy and Fung, Daniel and Guan, Cuntai and Lee, Tih-Shih and others},
  journal={NeuroImage: Clinical},
  volume={21},
  pages={101600},
  year={2019},
  publisher={Elsevier}
}

@article{rubinov2010complex,
  title={Complex network measures of brain connectivity: uses and interpretations},
  author={Rubinov, Mikail and Sporns, Olaf},
  journal={Neuroimage},
  volume={52},
  number={3},
  pages={1059--1069},
  year={2010},
  publisher={Elsevier}
}

@article{dey2012exploiting,
  title={Exploiting the brain's network structure in identifying ADHD subjects},
  author={Dey, Soumyabrata and Rao, A Ravishankar and Shah, Mubarak},
  journal={Frontiers in systems neuroscience},
  volume={6},
  pages={75},
  year={2012},
  publisher={Frontiers Media SA}
}

@article{yeo2011organization,
  title={The organization of the human cerebral cortex estimated by intrinsic functional connectivity},
  author={Yeo, BT Thomas and Krienen, Fenna M and Sepulcre, Jorge and Sabuncu, Mert R and Lashkari, Danial and Hollinshead, Marisa and Roffman, Joshua L and Smoller, Jordan W and Z{\"o}llei, Lilla and Polimeni, Jonathan R and others},
  journal={Journal of neurophysiology},
  year={2011},
  publisher={American Physiological Society Bethesda, MD}
}

@article{cai2018aberrant,
  title={Aberrant time-varying cross-network interactions in children with attention-deficit/hyperactivity disorder and the relation to attention deficits},
  author={Cai, Weidong and Chen, Tianwen and Szegletes, Luca and Supekar, Kaustubh and Menon, Vinod},
  journal={Biological Psychiatry: Cognitive Neuroscience and Neuroimaging},
  volume={3},
  number={3},
  pages={263--273},
  year={2018},
  publisher={Elsevier}
}

@article{hu2026fcn,
  title={FCN-LLM: Empower LLM for Brain Functional Connectivity Network Understanding via Graph-level Multi-task Instruction Tuning},
  author={Hu, Xingcan and Wang, Wei and Xiao, Li},
  journal={arXiv preprint arXiv:2603.01135},
  year={2026}
}

@article{li2026language,
  title={When Language Models Meet NeuroGraphs: Exploring Enhanced Agentic LLM Framework Towards Brain Network Analysis},
  author={Li, Jiaxing and Dong, Rui and Tang, Muyao and Kong, Youyong},
  journal={arXiv preprint arXiv:2607.22082},
  year={2026}
}

@inproceedings{zhang2024can,
  title={Can LLM graph reasoning generalize beyond pattern memorization?},
  author={Zhang, Yizhuo and Wang, Heng and Feng, Shangbin and Tan, Zhaoxuan and Han, Xiaochuang and He, Tianxing and Tsvetkov, Yulia},
  booktitle={Findings of the association for computational linguistics: EMNLP 2024},
  pages={2289--2305},
  year={2024}
}

@article{adhd2012adhd,
  title={The ADHD-200 consortium: a model to advance the translational potential of neuroimaging in clinical neuroscience},
  author={ADHD-200 consortium},
  journal={Frontiers in systems neuroscience},
  volume={6},
  pages={62},
  year={2012},
  publisher={Frontiers Research Foundation}
}

@article{schwarz2011negative,
  title={Negative edges and soft thresholding in complex network analysis of resting state functional connectivity data},
  author={Schwarz, Adam J and McGonigle, John},
  journal={Neuroimage},
  volume={55},
  number={3},
  pages={1132--1146},
  year={2011},
  publisher={Elsevier}
}

@article{bellec2017neuro,
  title={The neuro bureau ADHD-200 preprocessed repository},
  author={Bellec, Pierre and Chu, Carlton and Chouinard-Decorte, Francois and Benhajali, Yassine and Margulies, Daniel S and Craddock, R Cameron},
  journal={Neuroimage},
  volume={144},
  pages={275--286},
  year={2017},
  publisher={Elsevier}
}

@article{turpin2023language,
  title={Language models don't always say what they think: Unfaithful explanations in chain-of-thought prompting},
  author={Turpin, Miles and Michael, Julian and Perez, Ethan and Bowman, Samuel},
  journal={Advances in Neural Information Processing Systems},
  volume={36},
  pages={74952--74965},
  year={2023}
}

@article{mcnemar1947note,
  title={Note on the sampling error of the difference between correlated proportions or percentages},
  author={McNemar, Quinn},
  journal={Psychometrika},
  volume={12},
  number={2},
  pages={153--157},
  year={1947},
  publisher={Cambridge University Press \& Assessment}
}

@article{dietterich1998approximate,
  title={Approximate statistical tests for comparing supervised classification learning algorithms},
  author={Dietterich, Thomas G},
  journal={Neural computation},
  volume={10},
  number={7},
  pages={1895--1923},
  year={1998},
  publisher={MIT Press One Rogers Street, Cambridge, MA 02142-1209, USA journals-info~…}
}

@article{truhn2023large,
  title={Large language models should be used as scientific reasoning engines, not knowledge databases},
  author={Truhn, Daniel and Reis-Filho, Jorge S and Kather, Jakob Nikolas},
  journal={Nature medicine},
  volume={29},
  number={12},
  pages={2983--2984},
  year={2023},
  publisher={Nature Publishing Group US New York}
}

@article{luo2025large,
  title={Large language models surpass human experts in predicting neuroscience results},
  author={Luo, Xiaoliang and Rechardt, Akilles and Sun, Guangzhi and Nejad, Kevin K and Y{\'a}{\~n}ez, Felipe and Yilmaz, Bati and Lee, Kangjoo and Cohen, Alexandra O and Borghesani, Valentina and Pashkov, Anton and others},
  journal={Nature human behaviour},
  volume={9},
  number={2},
  pages={305--315},
  year={2025},
  publisher={Nature Publishing Group UK London}
}

@inproceedings{sclar2024quantifying,
  title={Quantifying Language Models' Sensitivity to Spurious Features in Prompt Design or: How I learned to start worrying about prompt formatting},
  author={Sclar, Melanie and Choi, Yejin and Tsvetkov, Yulia and Suhr, Alane},
  booktitle={International Conference on Learning Representations},
  volume={2024},
  pages={25055--25083},
  year={2024}
}

@inproceedings{nguyen2026robust,
  title={How Robust Are Large Language Models for Clinical Numeracy? An Empirical Study on Numerical Reasoning Abilities in Clinical Contexts},
  author={Nguyen, Minh-Vuong and Shiri, Fatemeh and Li, Zhuang and Verspoor, Karin},
  booktitle={Findings of the Association for Computational Linguistics: ACL 2026},
  pages={18113--18129},
  year={2026}
}

@inproceedings{schwartz2024numerologic,
  title={Numerologic: Number encoding for enhanced llms’ numerical reasoning},
  author={Schwartz, Eli and Choshen, Leshem and Shtok, Joseph and Doveh, Sivan and Karlinsky, Leonid and Arbelle, Assaf},
  booktitle={Proceedings of the 2024 Conference on Empirical Methods in Natural Language Processing},
  pages={206--212},
  year={2024}
}

@article{sporns2007identification,
  title={Identification and classification of hubs in brain networks},
  author={Sporns, Olaf and Honey, Christopher J and K{\"o}tter, Rolf},
  journal={PloS one},
  volume={2},
  number={10},
  pages={e1049},
  year={2007},
  publisher={Public Library of Science San Francisco, USA}
}

@article{wei2022chain,
  title={Chain-of-thought prompting elicits reasoning in large language models},
  author={Wei, Jason and Wang, Xuezhi and Schuurmans, Dale and Bosma, Maarten and Xia, Fei and Chi, Ed and Le, Quoc V and Zhou, Denny and others},
  journal={Advances in neural information processing systems},
  volume={35},
  pages={24824--24837},
  year={2022}
}

@article{biswal1995functional,
  title={Functional connectivity in the motor cortex of resting human brain using echo-planar MRI},
  author={Biswal, Bharat and Zerrin Yetkin, F and Haughton, Victor M and Hyde, James S},
  journal={Magnetic resonance in medicine},
  volume={34},
  number={4},
  pages={537--541},
  year={1995},
  publisher={Wiley Online Library}
}

@article{wang2023can,
  title={Can language models solve graph problems in natural language?},
  author={Wang, Heng and Feng, Shangbin and He, Tianxing and Tan, Zhaoxuan and Han, Xiaochuang and Tsvetkov, Yulia},
  journal={Advances in Neural Information Processing Systems},
  volume={36},
  pages={30840--30861},
  year={2023}
}

@inproceedings{petroni2019language,
  title={Language models as knowledge bases?},
  author={Petroni, Fabio and Rockt{\"a}schel, Tim and Riedel, Sebastian and Lewis, Patrick and Bakhtin, Anton and Wu, Yuxiang and Miller, Alexander},
  booktitle={Proceedings of the 2019 conference on empirical methods in natural language processing and the 9th international joint conference on natural language processing (EMNLP-IJCNLP)},
  pages={2463--2473},
  year={2019}
}

@article{singhal2025toward,
  title={Toward expert-level medical question answering with large language models},
  author={Singhal, Karan and Tu, Tao and Gottweis, Juraj and Sayres, Rory and Wulczyn, Ellery and Amin, Mohamed and Hou, Le and Clark, Kevin and Pfohl, Stephen R and Cole-Lewis, Heather and others},
  journal={Nature medicine},
  volume={31},
  number={3},
  pages={943--950},
  year={2025},
  publisher={Nature Publishing Group US New York}
}

@inproceedings{poerner2020bert,
  title={E-BERT: Efficient-yet-effective entity embeddings for BERT},
  author={Poerner, Nina and Waltinger, Ulli and Sch{\"u}tze, Hinrich},
  booktitle={Findings of the Association for Computational Linguistics: EMNLP 2020},
  pages={803--818},
  year={2020}
}

@article{grattafiori2024llama,
  title={The llama 3 herd of models},
  author={Grattafiori, Aaron and Dubey, Abhimanyu and Jauhri, Abhinav and Pandey, Abhinav and Kadian, Abhishek and Al-Dahle, Ahmad and Letman, Aiesha and Mathur, Akhil and Schelten, Alan and Vaughan, Alex and others},
  journal={arXiv preprint arXiv:2407.21783},
  year={2024}
}

@article{hui2024qwen2,
  title={Qwen2. 5-coder technical report},
  author={Hui, Binyuan and Yang, Jian and Cui, Zeyu and Yang, Jiaxi and Liu, Dayiheng and Zhang, Lei and Liu, Tianyu and Zhang, Jiajun and Yu, Bowen and Lu, Keming and others},
  journal={arXiv preprint arXiv:2409.12186},
  year={2024}
}

@article{yang2025qwen3,
  title={Qwen3 technical report},
  author={Yang, An and Li, Anfeng and Yang, Baosong and Zhang, Beichen and Hui, Binyuan and Zheng, Bo and Yu, Bowen and Gao, Chang and Huang, Chengen and Lv, Chenxu and others},
  journal={arXiv preprint arXiv:2505.09388},
  year={2025}
}

@article{guo2025deepseek,
  title={Deepseek-r1: Incentivizing reasoning capability in llms via reinforcement learning},
  author={Guo, Daya and Yang, Dejian and Zhang, Haowei and Song, Junxiao and Wang, Peiyi and Zhu, Qihao and Xu, Runxin and Zhang, Ruoyu and Ma, Shirong and Bi, Xiao and others},
  journal={arXiv preprint arXiv:2501.12948},
  year={2025}
}

@article{dettmers2023qlora,
  title={Qlora: Efficient finetuning of quantized llms},
  author={Dettmers, Tim and Pagnoni, Artidoro and Holtzman, Ari and Zettlemoyer, Luke},
  journal={Advances in neural information processing systems},
  volume={36},
  pages={10088--10115},
  year={2023}
}

@article{menon2011large,
  title={Large-scale brain networks and psychopathology: a unifying triple network model},
  author={Menon, Vinod},
  journal={Trends in cognitive sciences},
  volume={15},
  number={10},
  pages={483--506},
  year={2011},
  publisher={Elsevier}
}

@inproceedings{sakib2025spurious,
  title={Spurious correlations and beyond: Understanding and mitigating shortcut learning in sdoh extraction with large language models},
  author={Sakib, Fardin Ahsan and Zhu, Ziwei and Grace, Karen Trister and Yetisgen-Yildiz, Meliha and Uzuner, Ozlem},
  booktitle={Proceedings of the 63rd Annual Meeting of the Association for Computational Linguistics (Volume 2: Short Papers)},
  pages={1097--1106},
  year={2025}
}

@inproceedings{jacovi2020towards,
  title={Towards faithfully interpretable NLP systems: How should we define and evaluate faithfulness?},
  author={Jacovi, Alon and Goldberg, Yoav},
  booktitle={Proceedings of the 58th annual meeting of the association for computational linguistics},
  pages={4198--4205},
  year={2020}
}

@article{afolabi2026faithful,
  title={Faithful or Just Plausible? Evaluating the Faithfulness of Closed-Source LLMs in Medical Reasoning},
  author={Afolabi, Halimat and Afolabi, Zainab and Friel, Elizabeth and Roberts, Jude and Ji-Xu, Antonio and Chen, Lloyd and Ogbomo, Egheosa and Imevbore, Emiliomo and Eneje, Phil and Ouahidi, Wissal El and others},
  journal={arXiv preprint arXiv:2603.13988},
  year={2026}
}

@inproceedings{yao2014functional,
  title={Functional network disruption in attention deficit hyperactivity disorder},
  author={Yao, Zhijun and Hu, Bin and Xie, Yuanwei and Wang, Wei and Liu, Ruiyue and Liang, Chuanjiang and Su, Yun},
  booktitle={2014 IEEE International Conference on Bioinformatics and Biomedicine (BIBM)},
  pages={84--90},
  year={2014},
  organization={IEEE}
}

@article{xu2026deepseek,
  title={Deepseek-v4: Towards highly efficient million-token context intelligence},
  author={Xu, Anyi and Lin, Bangcai and Xue, Bing and Wang, Bingxuan and Xu, Bingzheng and Wu, Bochao and Zhang, Bowei and Lin, Chaofan and Dong, Chen and Ling, Chenchen and others},
  journal={arXiv preprint arXiv:2606.19348},
  year={2026}
}

@misc{openai2026gpt56luna,
author = {{OpenAI}},
title = {{GPT-5.6 Luna}},
year = {2026},
howpublished ={\url{https://developers.openai.com/api/docs/models/gpt-5.6-luna}},
note = {Accessed: 2026-09-21}
}

@article{dai2024revisiting,
  title={Revisiting the graph reasoning ability of large language models: Case studies in translation, connectivity and shortest path},
  author={Dai, Xinnan and Wen, Qihao and Shen, Yifei and Wen, Hongzhi and Li, Dongsheng and Tang, Jiliang and Shan, Caihua},
  journal={arXiv preprint arXiv:2408.09529},
  year={2024}
}

@inproceedings{zhang2024llm4dyg,
  title={LLM4DyG: Can large language models solve spatial-temporal problems on dynamic graphs?},
  author={Zhang, Zeyang and Wang, Xin and Zhang, Ziwei and Li, Haoyang and Qin, Yijian and Zhu, Wenwu},
  booktitle={Proceedings of the 30th ACM SIGKDD conference on knowledge discovery and data mining},
  pages={4350--4361},
  year={2024}
}

@inproceedings{wang2024unveiling,
  title={Unveiling factual recall behaviors of large language models through knowledge neurons},
  author={Wang, Yifei and Chen, Yuheng and Wen, Wanting and Sheng, Yu and Li, Linjing and Zeng, Daniel Dajun},
  booktitle={Proceedings of the 2024 Conference on Empirical Methods in Natural Language Processing},
  pages={7388--7402},
  year={2024}
}

@inproceedings{nishida2026revisiting,
  title={Revisiting Non-Verbatim Memorization in Large Language Models: The Role of Entity Surface Forms},
  author={Nishida, Yuto and Shikoda, Naoki and Kishinami, Yosuke and Fujii, Ryo and Morishita, Makoto and Kamigaito, Hidetaka and Watanabe, Taro},
  booktitle={Proceedings of the 64th Annual Meeting of the Association for Computational Linguistics (Volume 1: Long Papers)},
  pages={47046--47061},
  year={2026}
}

@article{guo2020shared,
  title={Shared and distinct resting functional connectivity in children and adults with attention-deficit/hyperactivity disorder},
  author={Guo, Xiaojie and Yao, Dongren and Cao, Qingjiu and Liu, Lu and Zhao, Qihua and Li, Hui and Huang, Fang and Wang, Yanfei and Qian, Qiujin and Wang, Yufeng and others},
  journal={Translational psychiatry},
  volume={10},
  number={1},
  pages={65},
  year={2020},
  publisher={Nature Publishing Group UK London}
}

@article{usuyama2025biomedical,
  title={Biomedical natural language processing in the era of large language models},
  author={Usuyama, Naoto and Wong, Cliff and Zhang, Sheng and Naumann, Tristan and Poon, Hoifung},
  journal={Annual Review of Biomedical Data Science},
  volume={8},
  number={1},
  pages={471--490},
  year={2025},
  publisher={Annual Reviews}
}

@article{gao2020pile,
  title={The pile: An 800gb dataset of diverse text for language modeling},
  author={Gao, Leo and Biderman, Stella and Black, Sid and Golding, Laurence and Hoppe, Travis and Foster, Charles and Phang, Jason and He, Horace and Thite, Anish and Nabeshima, Noa and others},
  journal={arXiv preprint arXiv:2101.00027},
  year={2020}
}

\end{document}